# Degradation of 2.4-kV Ga₂O₃ Schottky Barrier Diode at High Temperatures up to 500 °C

Hunter Ellis, Wei Jia, Imteaz Rahaman, Apostoli Hillas, Botong Li, Michael A. Scarpulla, Berardi Sensale Rodriguez, Kai Fu

*Abstract*— Ga₂O₃ Schottky barrier diodes featuring a field plate and a composite SiO₂/SiN$_x$ dielectric layer beneath the field plate were fabricated, achieving a breakdown voltage of 2.4 kV at room temperature. Electrical performance and degradation were analyzed via *I-V* and *C-V* measurements from 25 °C to 500 °C, revealing temperature-dependent transport, interface stability, and device stability. Upon returning to room temperature, the diodes exhibited nearly unchanged forward characteristics, while the breakdown voltage declined significantly from 2.4 kV to 700 V. This behavior indicates a temperature-induced reduction in the barrier height. Detailed analysis revealed that variable range hopping (VRH) dominated the leakage mechanism at moderate temperatures, while thermal emission (TE) became increasingly significant at temperatures exceeding 400 °C.

## I. Introduction

POWER electronics play a critical role in minimizing system power loss, but they encounter challenges in high-temperature environments such as deep oil drilling, planetary exploration, turbine engines, and geothermal energy systems [1], [2], [3], [4]. Wide bandgap and ultra-wide bandgap semiconductors present promising solutions to these challenges. Among them, *β*-Ga₂O₃ stands out due to its exceptional properties, including a wide bandgap of 4.85 eV, a high breakdown electric field of 8 MV/cm, and a melting point of 1800 °C [5]. Additionally, *β*-Ga₂O₃'s low intrinsic carrier density of $1.79 \times 10^{-23}$ cm⁻³ ensures that the intrinsic carrier concentration remains significantly lower than the doping levels, even at elevated temperatures [6]. These characteristics make *β*-Ga₂O₃ an excellent candidate for high-temperature applications [7], [8].

Schottky barrier diodes (SBD) are one of the fundamental and crucial power devices [9], [10], [11]. *β*-Ga₂O₃-based SBDs have been extensively studied by numerous research groups [12], [13], [14], [15], [16]. Recently, it has been observed that the Schottky contacts on *β*-Ga₂O₃ experience significant degradation in both forward and reverse performance at elevated temperatures [17], [18], [19], [20], [21]. For example, Ga₂O₃ SBDs have severely reduced breakdown voltages (BV) and increased leakage current at temperatures from 300 °C to 500 °C, with the leakage current rising by a factor of 10⁴ and the BV decreasing by more than ten times compared to room temperature (RT) [19], [20]. Despite its significance, this issue remains unexplored and insufficiently analyzed.

In this work, *β*-Ga₂O₃ Schottky barrier diodes are fabricated with a breakdown voltage of 2.4 kV at room temperature, and their degradation is investigated at high temperatures up to 500 °C. Key parameters such as ideality factor, barrier height, interface state density, leakage current, and breakdown voltage are analyzed through both *I-V* and *C-V* measurements. The underlying leakage mechanisms are also examined through detailed characterization.

## II. Device Fabrication and Measurements

The *β*-Ga₂O₃ diode structure was fabricated using a *β*-Ga₂O₃ wafer sourced from Novel Crystal Technology, Inc. The wafer featured a 10 μm thick drift layer with a doping concentration of $1 \times 10^{16}$ cm⁻³, grown on a (001) Sn doped n⁺-*β*-Ga₂O₃ substrate. The fabrication process is shown in Fig. 1(a). A 35 μm thick photoresist was employed for etching *β*-Ga₂O₃, resulting in a 1 μm deep trench isolation and lifting off the following dielectric. The dielectric layer was deposited by alternating the sputtering of 40 nm layers of SiO₂ and SiN$_X$, repeated until the 1 μm deep trench was fully filled. Subsequently, Pt (50 nm)/Au (250 nm) alloy was deposited to form a Schottky contact on the *β*-Ga₂O₃ and a field plate (FP) on the dielectric simultaneously. The diode has a diameter of 200 μm, with a FP length of 45 μm. Finally, Ti (50 nm)/Au (250 nm) was deposited on the bottom surface to form the ohmic

This work is supported in part by the University of Utah start-up fund and PIVOT Energy Accelerator Grant U-7352FuEnergyAccelerator2023. (*Corresponding author: Kai Fu*). Hunter Ellis and Wei Jia contributed equally, or if there are co-first authors.
The authors are with the Electrical and Computer Engineering Department, the John and Marcia Price College of Engineering, the University of Utah, Salt Lake City, UT 84112, USA.

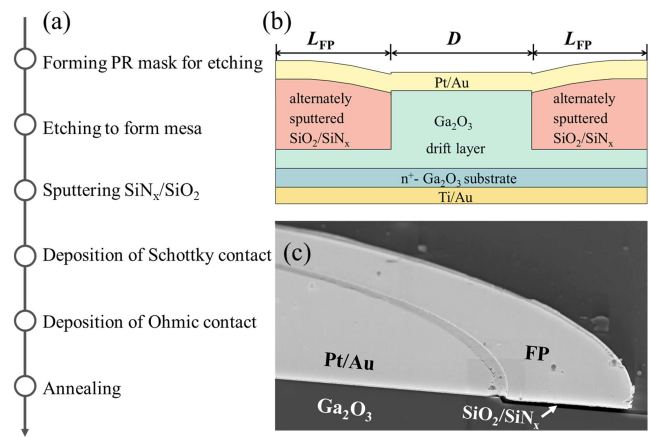

Fig. 1. (a) Process flow. (b) Schematic of the *β*-Ga₂O₃ SBD structure. (c) SEM image showing the cross-section of the SBD.

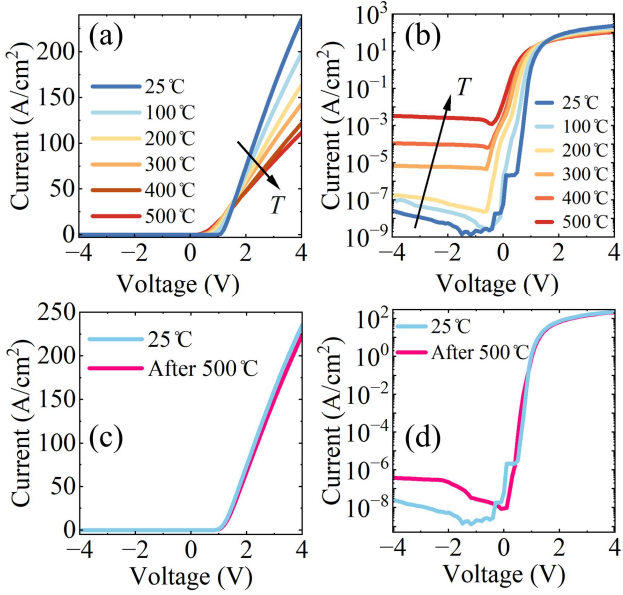

Fig. 2. (a) Linear and (b) semi-log forward $I$-$V$ curves from 25 °C to 500 °C. Comparison of $I$-$V$ curves before and after 500 °C in (c) linear and (b) semi-log scale.

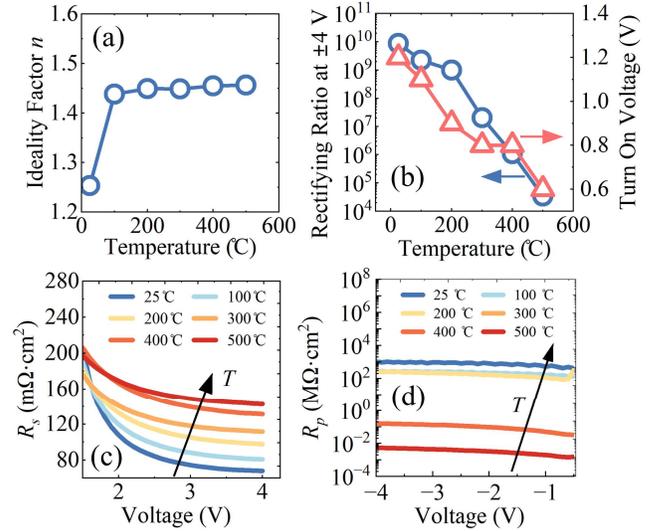

Fig. 3. (a) Ideality factor of the $Ga_2O_3$ diode at different temperatures. (b) Rectifying ratio at ±4 V and the turn-on voltage from 25 °C to 500 °C. (c) Derived series ($R_s$) and (d) parallel ($R_p$) resistance as a function of temperature.

contact. An SEM image of the device structure is shown in Fig. 1(b). The forward $I$-$V$ and $C$-$V$ measurements were characterized using a 4200A-SCS parameter analyzer, while the breakdown voltage was measured up to 1000 V using a Keithley 2470 SourceMeter. For higher voltages, up to 10 kV, a Matsusada AU30N5 was used, with 3M™ Fluorinert™ Electronic Liquid FC-40 applied to prevent air breakdown.

III. FORWARD PERFORMANCE AT HIGH TEMPERATURES

Figure 2(a) and (b) present the forward $I$-$V$ curves of the $β$-$Ga_2O_3$ diode in linear and semi-log scales, respectively, measured from 25 °C to 500 °C. The reverse leakage current increased significantly with temperature (from $10^{-8}$ to $10^{-3}$ A/cm$^2$), while the on-current declined. The devices were held at each temperature for 1 hour. Figure 2(c) and (d) illustrate the recovery behavior of the diode after testing up to 500 °C. After 5 hours of high-temperature storage, the forward current fully recovered, but the reverse leakage current exhibited a minor and permanent increase from $10^{-8}$ to $10^{-7}$ A/cm$^2$.

Figure 3(a) presents the ideality factor, which increases with temperature, likely due to a nonhomogeneous Schottky contact with temperature-dependent interface state densities [22], [23], [24], [25], [26]. Figure 3(b) shows the rectifying ratio at ±4 V, which decreases with temperature as the off-current increases and the on-current decreases. The turn-on voltage (defined at 100 μA) also decreases with rising temperature, primarily due to increased leakage current. Figure 3(c) and (d) depict the temperature-dependent equivalent series ($R_s$) and parallel resistance ($R_p$), respectively. The $R_s$ increases with temperature due to reduced mobility at higher temperatures [19] since the carrier concentration does not change much with the temperature (Fig. 4(b)), while $R_p$ decreases, likely driven by faster trap de-trapping rates and elevated thermal emission currents [27].

Figure 4(a) displays the $C$-$V$ measurements at 1 MHz, while Fig. 4(b) illustrates the ionized donor's depth profile from 25 °C to 500 °C, with an average doping concentration of $1×10^{16}$ cm$^{-3}$ calculated from the $C$-$V$ data. Figure 4(c) presents the $C^{-2}$–$V$ plots across the same temperature range, with the inset depicting a equivalent circuit model of the diode's capacitance [28]. This model incorporates the depletion capacitance beneath the Schottky contact ($C_D$), in parallel with the FP capacitance from the dielectric ($C_{FP}$) and the depletion capacitance beneath the FP in the $β$-$Ga_2O_3$ ($C_{FD}$), describing the total system capacitance as given by Eq. (1).

$$C = C_d+(C_{FP}×C_{FD}/(C_{FP}+C_{FD})) \quad (1)$$

$$C = C_d+C_{FP} \quad (2)$$

$$C = b_1(V−φ_{bi})^{-1/2}+b_2 \quad (3)$$

$$C_D = (qε_sN_D/2(V-φ_{bi}))^{1/2} \quad (4)$$

Given that $C_{FP}$ is significantly smaller than $C_{FD}$ due to the dielectric's thickness of 1 μm, Eq. (1) can be approximated by Eq. (2). The data was subsequently fitted by minimizing the mean squared error in Eq. (3) for $b_1$ and $b_2$. Here, $b_1$ represents the coefficient for the voltage-dependent capacitance $C_D$, $b_2$ corresponds to $C_{FP}$, $V$ denotes the applied voltage, and $φ_{bi}$ is the built-in potential [28]. The equation for $C_D$ is given in Eq. (4), where $N_D$ is the donor concentration, $q$ is the charge of an electron, and $ε_s$ is the permittivity of $β$-$Ga_2O_3$. With the increase of reverse bias, the $C^{-2}$ data deviates from linearity due to the influence of the additional FP capacitance [29]. Fig. 4(d) shows a magnified view of $C^{-2}$–$V$ data from Fig. 4(c), revealing an increase in capacitance with temperature. The trend suggests a reduction in depletion width, which can be attributed to a decrease in barrier height.

To examine the barrier height variation, it was derived from both $I$-$V$ and $C$-$V$ measurements for comparison. The saturation current $I_0$, a key parameter for determining the barrier height from the $I$-$V$ curve, was obtained from the intercept of the extrapolated $I$-$V$ data, as shown in Fig. 5(a). The derived $I_0$ values across the temperature range from room temperature to 500 °C are presented in Fig. 5(b). Based on the expression for $I_0$ given in Eq. (5), $I_0$ increases with temperature, where $A^*$ is the Richardson constant for $\beta$-$Ga_2O_3$ (41 A·cm$^{-2}$·k$^{-2}$) [30], $k$ is the Boltzmann constant, $\varphi_b$ is the barrier height, $A$ is the area of the diode, and $T$ is the temperature. Interestingly, we observed that $I_0$ increases linearly with the temperature (in °C) on log-log scale. The empirical relationship is expressed in Eq. (6).

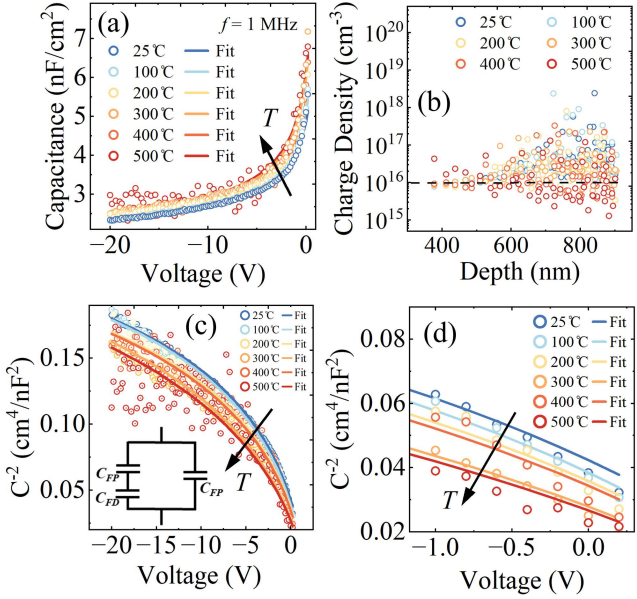

Fig. 4. (a) $C$-$V$ curves from 25 °C to 500 °C. (b) Charge density in the $Ga_2O_3$. (c) $C^{-2}$-$V$ curves at different temperatures, with inset showing the equivalent circuit considering the capacitance under the field plate (FP). (d) $C^{-2}$-$V$ at different temperatures showing the shift of built-in potential.

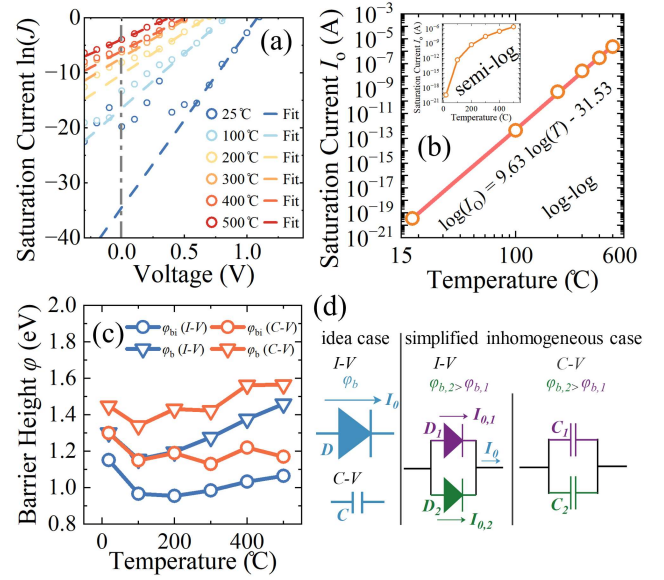

Fig. 5. (a) linear fitting for forward $I$-$V$ data. (b) $I_0$ from 25 °C to 500 °C. (c) Schottky barrier height $\varphi_b$ and built-in potential $\varphi_{bi}$ as a function of temperature derived from the $I$-$V$ and $C$-$V$ curves. (d) Schematics of ideal Schottky diode with leakage current $I_0$, simplified inhomogeneous Schottky diode with two distinct barriers represented by $D_1$ and $D_2$, and equivalent capacitive circuits.

$$I_0 = AA^*kT^2\exp(q\varphi_b/kT) \quad (5)$$

$$\log(I_0) = 9.63 \times \log(T) - 31.53 \quad (6)$$

Figure 5(c) compares the barrier height ($\varphi_b$) and the built-in potential ($\varphi_{bi}$) derived from $I$-$V$ and $C$-$V$ measurements, where their relationship is described in Eq. (7), with $N_C$ representing the effective density of states of conduction band [29]. The observed discrepancy between $\varphi_b$ and $\varphi_{bi}$ values from $I$-$V$ and $C$-$V$ methods suggests an inhomogeneous Schottky barrier [22], [30], usually caused by interface defects, metal grain boundaries, and/or contact non-uniformities [31], [32]. A simplified model, illustrated in Fig. 5(d), represents the diode as having two distinct barrier heights and corresponding $I_0$. This diode can be extended to a simplified model, as shown in Fig 5(d), to represent a diode with two distinct barrier heights. This model can be extended to include multiple parallel diodes or a continuous distribution of barrier heights, capturing the complexity of real-world barrier inhomogeneities [22], [28], [33], [34], [35], [36], [37].

$$\varphi_b = \varphi_{bi} + kT\ln(N_C/N_D) \quad (7)$$

The barrier height calculated from $I$-$V$ measurements is primarily influenced by the regions with lower barrier heights, as it is derived from $I_0$, which is exponentially sensitive to $\varphi_b$.

In the simplified case of two distinct barriers, as depicted in Fig. 5(d), the total $I_0$ is given by Eq. (8), where $I_{0,1}$ and $I_{0,2}$ represent the saturation currents for diodes $D_1$ and $D_2$, respectively. Given the exponential dependence of $I_0$ on $\varphi_b$ as shown in Eq. (5), even minor variations in $\varphi_b$ can lead to significant changes in $I_0$, highlighting the dominance of lower barriers in determining the $I$-$V$-derived barrier height.

$$I_0 = I_{0,1} + I_{0,2} \quad (8)$$

For example, let $\varphi_{b,1}$, the barrier height for $D_1$, be 1 eV, and $\varphi_{b,2}$, the barrier height for $D_2$, be 1.15 eV. These values result in $I_{0,1} = 5.1 \times 10^{-15}$ A/cm$^3$ and $I_{0,2} = 1.6 \times 10^{-17}$ A/cm$^3$. Since $I_{0,1}$ is nearly two orders of magnitude larger than $I_{0,2}$, Equation (7) can be approximated by considering only $I_{0,1}$.

In contrast, the barrier heights obtained from $C$-$V$ measurements reflect the combined capacitance contributions of both diodes, as modeled in the parallel capacitor circuit in Fig. 5(d). The total capacitance $C_{tot}$, calculated using Eq. (9), is the area-weighted sum of the individual capacitances ($C_1$ and $C_2$), where $A_1$ and $A_2$ are the respective contact areas.

$$C_{tot} = (A_1C_1 + A_2C_2)/A \quad (9)$$

As defined in Eq. (r), the capacitance of an SBD has a square-root dependence on $\varphi_{bi}$, such that small variations in $\varphi_{bi}$ lead to only minor changes in the capacitance. For the barrier heights considered in the previous example, the corresponding capacitances are $C_1 = 28$ nF and $C_2 = 26$ nF based on the $\varphi_{bi}$. The relatively small difference between $C_1$ and $C_2$ suggests that barrier heights derived from $C$-$V$ measurements are less sensitive to lower barrier regions in inhomogeneous systems compared to those obtained from $I$-$V$ measurements.

## IV. INTERFACE DENSITY

Figure 6(a) displays the room-temperature $C$-$V$ data measured across frequencies from 1 MHz to 7 kHz, highlighting the frequency-dependent behavior of interface states in the SBD. At frequencies above 100 kHz, the interface states have time constants too large to contribute significantly, resulting in minimal capacitance variation between 100 kHz and 1 MHz. In contrast, at lower frequencies, such as 7 kHz, all interface states can respond, causing the capacitance to stabilize at the same value observed at 10 kHz. Figure 6(b) depict the interface state density ($N_{ss}$), which increases with temperature, correlating with the observed rise in leakage current. $N_{ss}$ is calculated using Eq. (10), where $C_{LF}(V)$ and $C_{HF}(V)$ represent the low-frequency (7 kHz) and high-frequency (1 MHz) capacitances, respectively [38], [39].

$$N_{ss} = (q\varepsilon_s N_D/2\varphi_{bi})^{1/2} \times (C_{LF}(V) - C_{HF}(V)) \times q C_{HF}(V) \quad (10)$$

The energy level of the interface state ($E_{ss}$) was calculated using Eq. (11), where $E_c$ is the conduction band energy, and $\varphi_e$ is the effective barrier height. $\varphi_e$ is defined in Eq. (12) and derived from barrier height ($\varphi_{b,iv}$) obtained from $I$-$V$ measurements [40].

$$E_c - E_{ss} = q(\varphi_e - V) \quad (11)$$

$$\varphi_e = \varphi_{b,iv} + (1 - n^{-1})V \quad (12)$$

Figure 6(c) illustrates a band diagram representing $N_{ss}$, which also highlights $\varphi_o$, the neutral level, which serves as the boundary: interface states above $\varphi_o$ exhibit acceptor-like behavior, while those below $\varphi_o$ exhibit donor-like behavior.

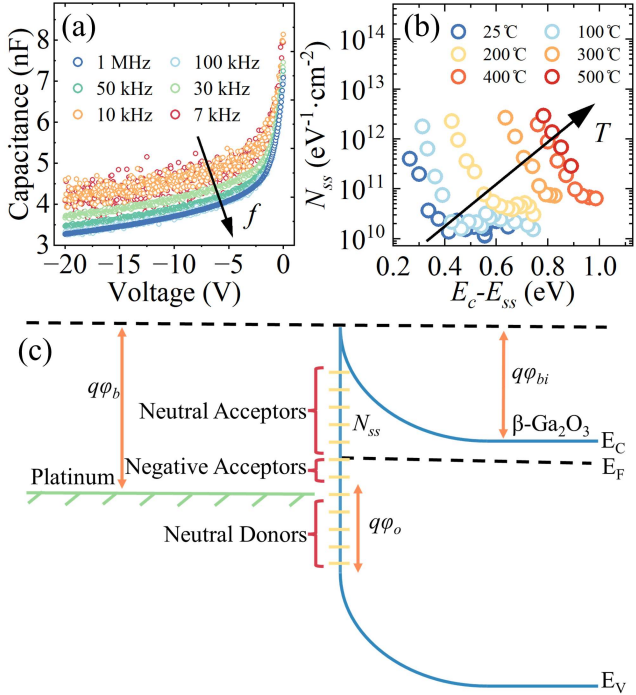

Fig. 6. (a) $C$-$V$ curves from 1 MHz to 7 kHz. (b) Interface state density at the interface of the Schottky contact and the β-Ga$_2$O$_3$ from 25 °C to 500 °C. (c) Schematic of energy band diagram showing interface states.

## V. REVERSE PERFORMANCE AT HIGH TEMPERATURES

Figure 7(a) shows the reverse bias $I$-$V$ curve at 25 °C with a breakdown voltage of around 2.4 kV, corresponding to an electric field strength of around 3.3 MV/cm. The inset displays a cleaved device post-breakdown. Figure 7(b) depicts reverse bias $I$-$V$ curves from 25°C to 500°C, revealing a significant increase in leakage current with temperature, particularly beyond 200°C. The breakdown voltage, initially above 1 kV at 25 °C and 100 °C, dropped to 700 V at 200 °C and fell below 100 V at higher temperatures. Figure 7(c) compares the reverse $I$-$V$ characteristics before and after 500°C testing, revealing a permanent reduction in breakdown voltage to 650 V. The degradation is expected to be attributed to increased interface states, reduced barrier height, and/or potential contact diffusion into the diode during high-temperature exposure [18], [41].

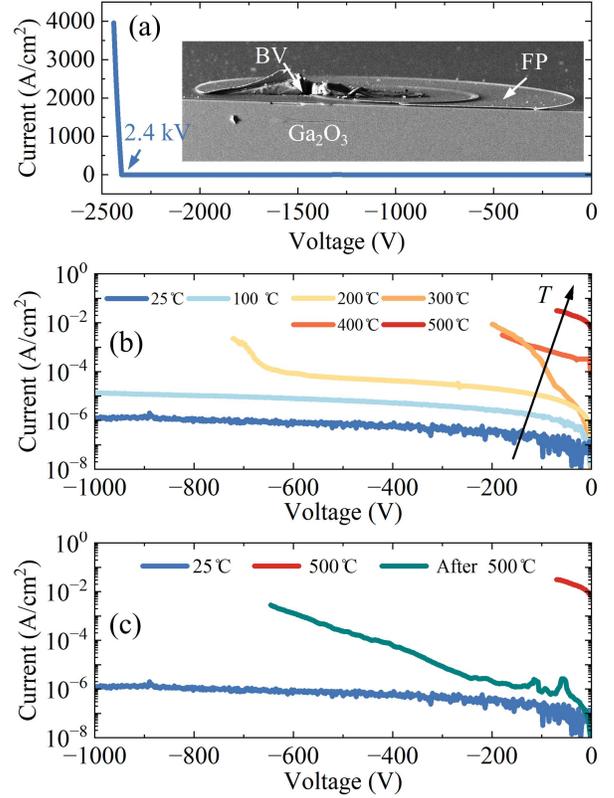

Fig. 7. (a) Reverse breakdown $I$-$V$ curve of the Ga$_2$O$_3$ SBD at room temperature. The inset shows the cross-section of a diode after breakdown. (b) Reverse $I$-$V$ curves of the β-Ga$_2$O$_3$ SBD from 25 °C to 500 °C. (c) Comparison of the reverse $I$-$V$ curves of the Ga$_2$O$_3$ SBD at 25 °C, at 500 °C, and after 500 °C.

Analysis of the reverse $I$-$V$ data reveals that leakage current scales linearly with area, as shown in Fig. 8(a), indicating that surface leakage is negligible. From 25°C to 200°C, the leakage mechanism follows the variable range hopping (VRH) model. Fitting results for the VRH process are presented in Fig. 8(b) and 8(c) for voltage ranges of 24–62 V and 100–700 V, respectively. The VRH current density ($J_{VRH}$) was calculated using Eq. (6), where $T_c$ is the characteristic temperature [42], [43].

$$J_{VRH} \propto \exp(-(T_c/T)^{1/4}) \quad (13)$$

With the increase of temperature, the leakage due to thermal emission (TE) becomes significant. Both TE and VRH account for the leakage. Figure 8(d) shows the VRH fitting for the data taken from 300 °C to 500 °C with the TE current subtracted. We have found that the $T_c$ in the VRH model increases with temperature under reverse bias, suggesting changes in material disorder or carrier transport dynamics. This could be attributed to: (a) activation of deeper trap states at higher temperatures, increasing the effective density of localized states and broadening energy levels involved in hopping; (b) enhanced carrier localization due to thermally activated occupancy of more localized states under reverse bias, reducing hopping distances. Figure 8(e) compares the measured current density with the simulated TE, VRH, and the combined contributions. TE modeled using Eq. (14), where $J_{TE}$ is TE the current density [22], and $E$ is the electric field at metal-$\beta$-Ga$_2$O$_3$ interface [44]. Figure 8(f) illustrates the calculated TE leakage current at 500 °C for a $\beta$-Ga$_2$O$_3$ diode with varying barrier heights. The results show that a high barrier height (e.g., > 2 eV), is crucial for suppressing current leakage at elevated temperatures.

$$J_{TE} \propto A^* T^2 \exp(-q\varphi_{b,iv} - (qE/4\pi\varepsilon_s)^{1/2}))(kT)^{-1}) V_o \quad (14)$$

$$V_o = (\exp(-qV/kT)) - 1 \quad (15)$$

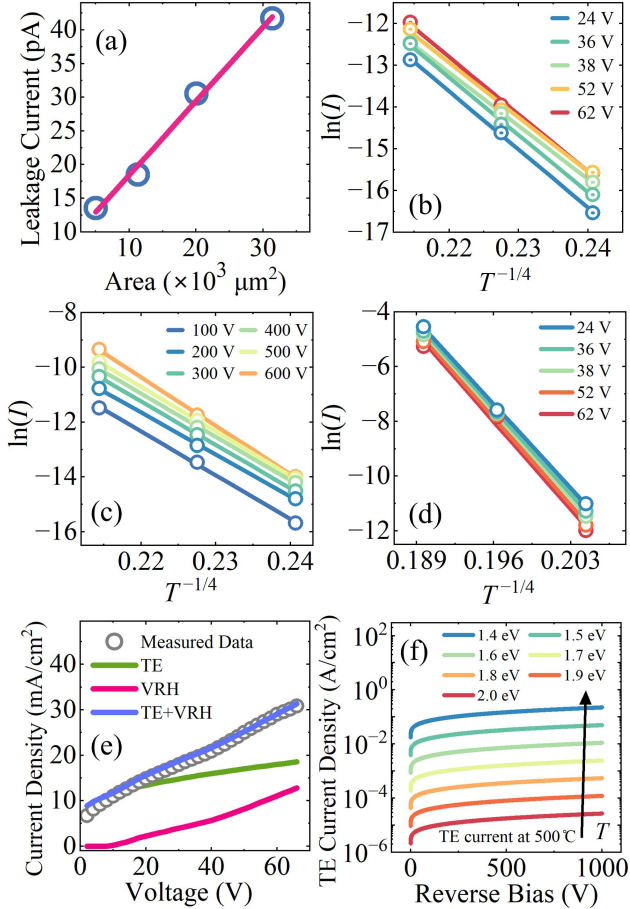

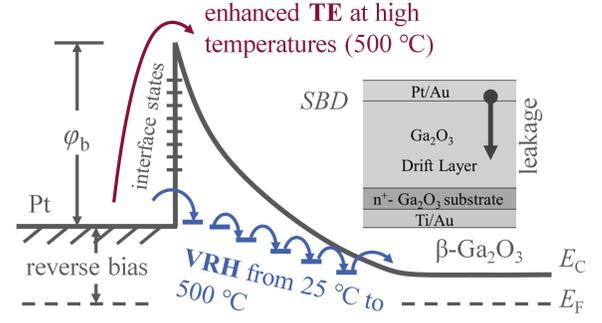

Fig. 9. Schematic of leakage mechanisms at high temperatures.

Figure 9 presents the band diagram of a $\beta$-Ga$_2$O$_3$ SBD to indicate the dominant leakage mechanisms across different temperatures. The VRH model prevails at intermediate temperatures, while the TE becomes increasingly significant as the temperature approaches 500°C.

## VI. CONCLUSION

A $\beta$-Ga$_2$O$_3$ SBD with a breakdown voltage of 2.4 kV was fabricated and analyzed via *I-V* and *C-V* measurements over 25°C to 500°C. Results revealed an inhomogeneous Schottky interface with increasing interface state densities and discrepancies between *C-V* and *I-V* barrier heights, contributing to elevated leakage currents and reduced breakdown voltage at higher temperatures. The dominant leakage mechanism transitioned from VRH to TE above 400°C. After cooling, the device retained forward characteristics, but the breakdown voltage decreased to 700 V.

Improving high-temperature performance could involve optimizing surface treatments, contact metallurgy, and employing metals with higher, stable work functions to mitigate TE leakage. Alternative designs, such as Schottky diodes made with PbCoO$_2$ and platinum group metal oxide contacts, have demonstrated resilience to high temperatures with low reverse and forward biases [10], [17], [45], [46]. Besides, devices employing pn heterojunctions, like NiO/$\beta$-Ga$_2$O$_3$ diodes, have shown improved performance, maintaining breakdown voltages above 1 kV even at high temperatures [20], [47], [48], [49], attributed to higher built-in voltages, larger depletion regions, and reduced defect densities [28].

## ACKNOWLEDGMENT

The device fabrication was performed at the Nanofab at the University of Utah.

Fig. 8. (a) Leakage current vs device contact area. Fitting curves based on VRH transportation at (b) low-temperature and low-voltage, (c) low-temperature and high-voltage, and (d) high-temperature and low-voltage. (e) Comparison of measured leakage current and calculated leakage current based on TE and VRH transport models. (f) Simulated reverse *I-V* with the barrier height varying from 1.4 eV to 2.0 eV.